\RequirePackage{amsthm}
\documentclass[pdflatex,sn-mathphys,Numbered]{sn-jnl}% Math and Physical Sciences Reference Style
\usepackage{graphicx}%
\usepackage{multirow}%
\usepackage{amsmath,amssymb,amsfonts}%
\usepackage{amsthm}%
\usepackage{mathrsfs}%
\usepackage[title]{appendix}%
\usepackage{xcolor}%
\usepackage{textcomp}%
\usepackage{manyfoot}%
\usepackage{booktabs}%
\usepackage{algorithm}%
\usepackage{algorithmicx}%
\usepackage{algpseudocode}%
\usepackage{listings}%
\usepackage{subfigure}%subfigures
\usepackage{float}%floating figures
\usepackage{dcolumn}% Align table columns on decimal point
\usepackage{bm}% bold math
\theoremstyle{thmstyleone}%
\theoremstyle{thmstyletwo}%

\theoremstyle{thmstylethree}%

\begin{document}

\title{Development of neutron beamline for laser-driven neutron resonance spectroscopy}


\author[1]{\fnm{Zechen} \sur{Lan}}\email{lan-zc@ile.osaka-u.ac.jp}

\author[1]{\fnm{Yasunobu} \sur{Arikawa}}\email{arikawa.yasunobu.ile@osaka-u.ac.jp}

\author[1]{\fnm{Alessio} \sur{Morace}}\email{morace.alessio.ile@osaka-u.ac.jp}

\author[2]{\fnm{Yuki} \sur{Abe}}\email{abe.yuki@eei.eng.osaka-u.ac.jp}

\author[3]{\fnm{S. Reza} \sur{Mirfayzi}}\email{abe.yuki@eei.eng.osaka-u.ac.jp}

\author[1]{\fnm{Tianyun} \sur{Wei}}\email{wei-t@osaka-u.ac.jp}

\author[4]{\fnm{Takehito} \sur{Hayakawa}}\email{hayakawa.takehito@qst.go.jp}

\author*[1]{\fnm{Akifumi} \sur{Yogo}}\email{yogo.akifumi.ile@osaka-u.ac.jp}

\affil*[1]{\orgdiv{Institute of laser engineering}, \orgname{Osaka University}, \orgaddress{ \country{Japan}}}

\affil[2]{\orgdiv{Graduate school of engineering}, \orgname{Osaka University}, \orgaddress{ \country{Japan}}}

\affil[3]{\orgname{Tokamak Energy Ltd}, \orgaddress{\country{United Kingdom}}}

\affil[4]{\orgname{National Institute for Quantum Science and Technology}, \orgaddress{ \country{Japan}}}

\abstract{Recent progress of laser science provides laser-driven neutron source (LDNS), which has remarkable features such as the short pulse width.
One of the key techniques  to be developed for more efficient use of the LDNS is neutron collimation tubes to increase the number of neutrons arriving at a detector in the time-of-flight method.
However, when a tube with a thick wall is used as a collimator the neutron collection efficiency at the detector increases but
the time resolution becomes wider because of multiple scattering inside of the tube. 
In the present study, we have developed a collimation tube made of Ni-0, which is optimized for
the increased neutron collection efficiency and a reasonable time resolution.
This collimator has been demonstrated experimentally using neutron resonance spectroscopy with neutrons provided from LFEX laser.}




\maketitle

\section{Introduction}

Laser-driven neutron source (LDNS) \cite{roth2013bright,kar2016beamed,alejo2017high,kleinschmidt2018intense,zimmer2022demonstration,yogo2023advances} is triggered by laser-plasma interactions, which have realized acceleration of ions up to the energies higher than the threshold of neutron-producing nuclear reactions such as d(p,~np)p, d(d, n)$^{3}$He, $^{9}$Be(p, n)$^{9}$B, $^{9}$Be(d, xn), and $^{7}$Li(p,~n)$^{7}$Be in the second target.
Compared to nuclear reactors or accelerator-based neutron facilities, LDNS has several distinctive features: i) the compactness of the neutron generation source size ($\sim$1 cm), ii) generation of pulsed neutrons shorter than 1~ns at the source, and iii) the high penetrating ability of the laser light, which extends the use of neutrons to the usable area of accelerator-based neutron sources.
The LDNS predominantly generates fast neutrons in the energy range of MeV, which have been applied to various subjects such as radiography of thick materials \cite{arikawa2023demonstration}, the study of nuclear physics \cite{gunther2022forward}, and production of medical raidoisotopes \cite{mori2023feasibility}.
In general, the cross sections of (n, $\gamma$) neutron capture reactions increase with decreasing the neutron energy (well known as the $1/v$ law), according to the Breit-Wigner formula in the low-energy region \cite{breit1936capture}.
Recently, the LDNS has provided neutrons in the energy of thermal $\sim$25 meV \cite{mirfayzi2020miniature} and epithermal $\sim$1-100~eV \cite{mirfayzi2019ultra} by locating a neutron moderator at the vicinity of the LDNS. These low-energy neutrons could be used for various applications such as isotope selective radiography and neutron resonance spectroscopy.
\\
Neutron energies are in general measured using a time-of-flight (TOF) method with a beamline with a length from a few meters to several tens of meters in accelerator-based neutron facilities in  \cite{tremsin2014neutron,schillebeeckx2012neutron} in addition to the LDNS.
Recently, a neutron resonance spectroscopy has been performed with epithermal neutrons generated from LDNS. It is noted that in these measurements, a beamline as short as 1.8 m was used \cite{yogo2023laser,zimmer2022demonstration}.
One of the key technologies for the neutron resonance spectroscopy using the LDNS is the small neutron moderator to minimize the statistical broadening of the neutron pulse during the moderation process.
However, the extremely short beamline for the LDNS is inevitably affected by backgrounds of flush x-rays generated directly from the laser-induced plasma and neutrons that are  scattered from the laser chamber and other constructions. 
The x-rays generally produce signals much higher than the epithermal neutron signals to be measured. 
To reduce the x-ray signals, time-gated photomultiplier tubes used for scintillation detectors have been developed, but it cannot remove the delayed neutron backgrounds.
The background neutrons arrive at the TOF detector in a long time range and thus it is difficult to separate these two neutron signals using time gate.
One of the approaches to decreases neutron backgrounds is development of a neutron collimator and a shield designed for the short beamline of the LDNS. 

In this paper, we report the development of a collimation tube made of nickel with natural abundance (Ni-0), which is optimized to enhance the neutron collection efficiency only with small broadening of the time resolution. We also report a boron-doped polyethylene shield for the neutron detector.
The effectiveness of the collimator and shield have been confirmed in a NRS experiment at LFEX laser facility of ILE, Osaka University. 
We show that the signal-to-noise (S/N) ratio of neutron resonance peaks are improved by a factor of 2.

\section{Development of collimator for LDNS}
 
%The developments of an nickle epithermal neutron collimator and  are introduced in this section.
\subsection{Development of Nickle epithermal neutron reflector} 
\begin{figure}[htbp]
	\subfigbottomskip=0pt
	\subfigcapskip=0pt 
	\begin{center}
		\includegraphics[width=0.4\textwidth]{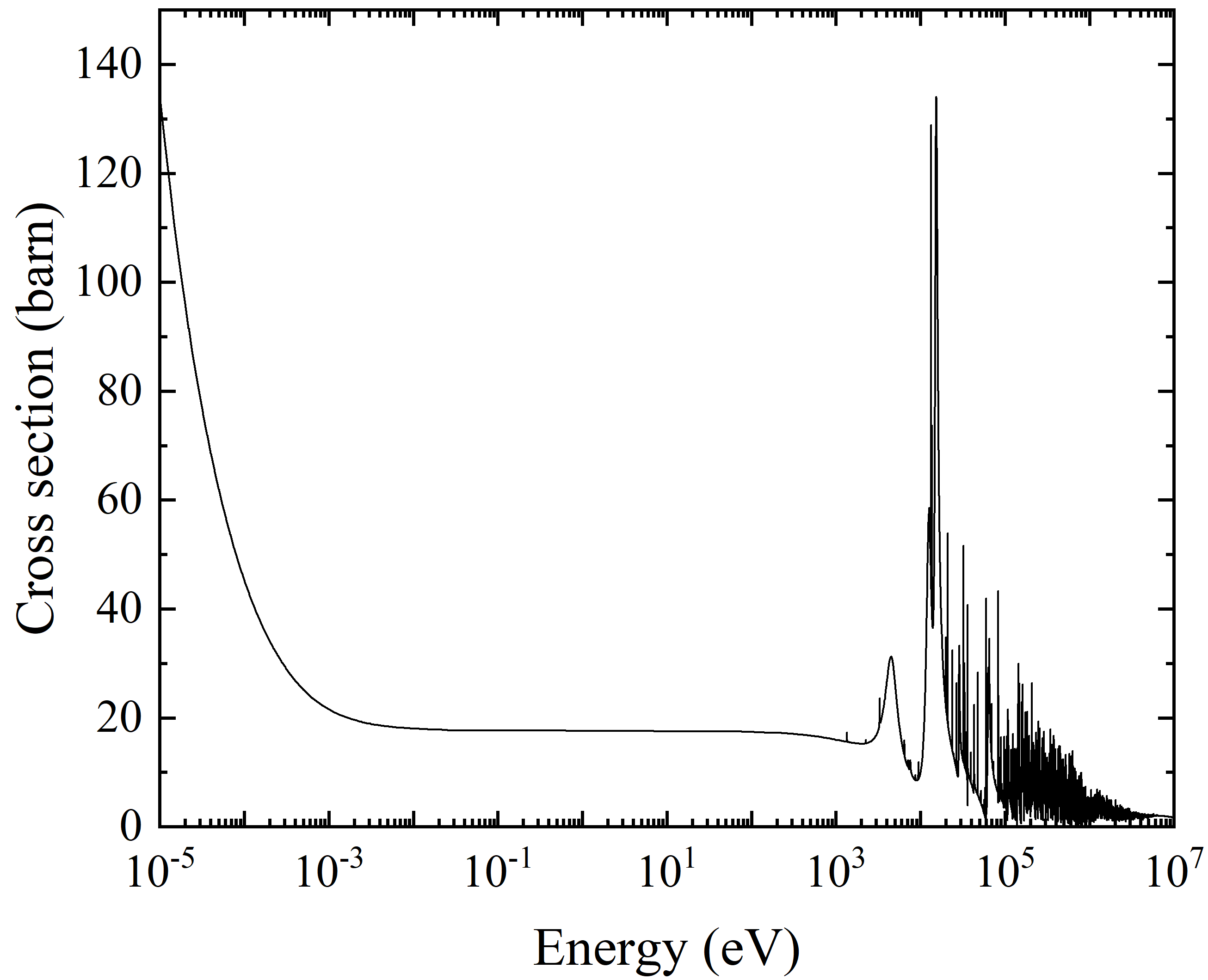}%
	\end{center}
	\caption{\label{F.NiElasC}The cross section of neutron elastic scattering in Ni-0. The evaluated nuclear data is obtained from JENDL4.0 database \cite{shibata2011jendl}.}
\end{figure}

\begin{figure}[htbp]
	\subfigbottomskip=0pt
	\subfigcapskip=0pt 
	\begin{center}
		\subfigure[\label{F.NiColli(a)}]{
			\includegraphics[width=0.5\textwidth]{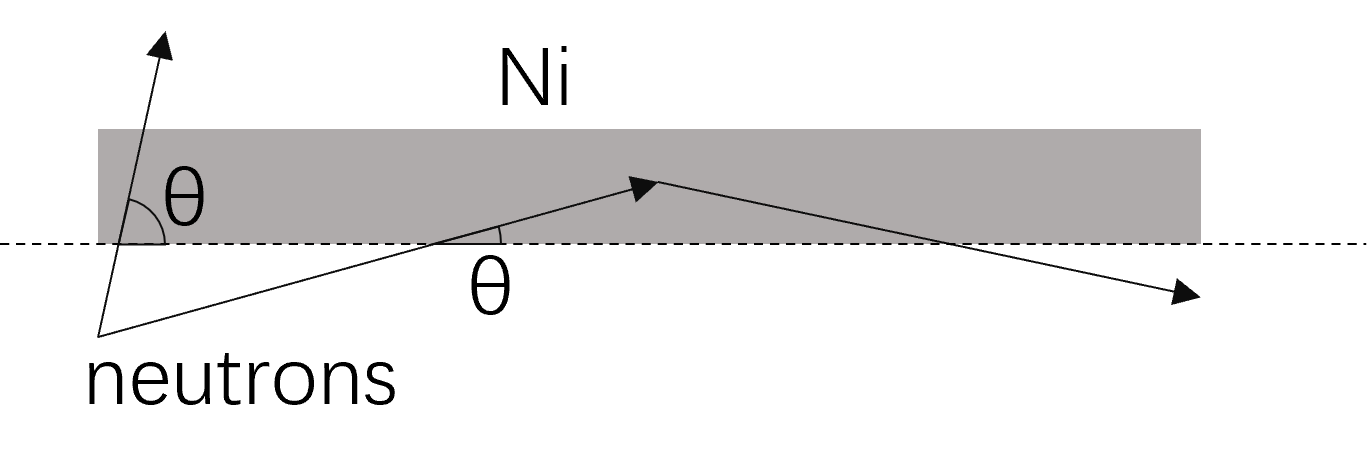}%
		}
		\\
		\subfigure[\label{F.NiColli(b)}]{
			\includegraphics[width=0.4\textwidth]{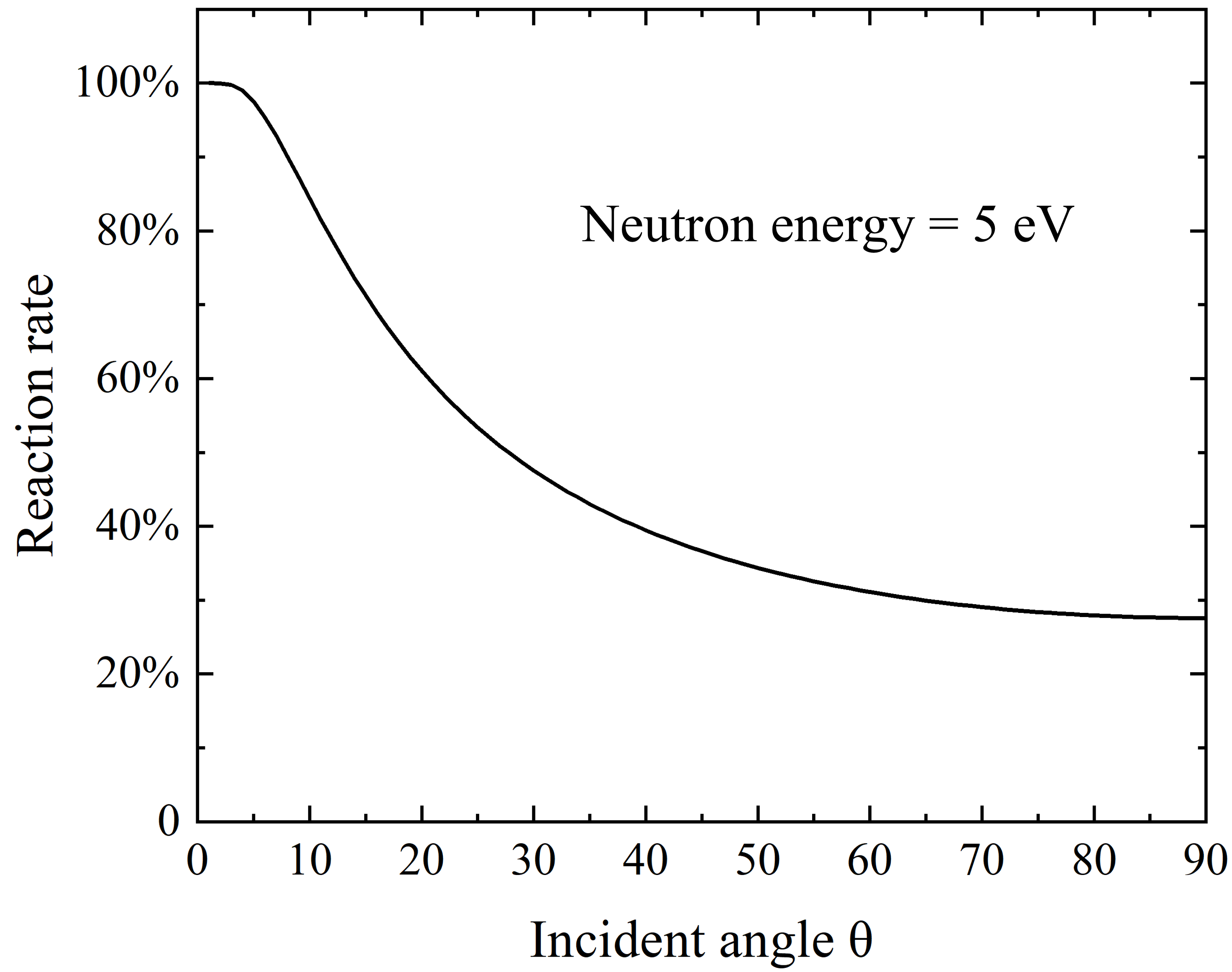}%
		}
	\end{center}
	\caption{\label{F.NiColli}(a) The schematic diagram of nickle collimator. (b) The reaction rate of 5eV neutrons elastic scattering changes with incident angles.} 
\end{figure}

In previous neutron experiments, neutron supermirrors and reflectors made of nickle have been widely used. The 'reflector' and 'mirror' refer to the elastic scattering of neutrons on atomic nuclei rather than the optical reflection. 
%
%In previous works for neutron reflectors, the interests are most focused on the critical angle that depends on the neutron wavelength. 
The nickle has relatively large cross sections of neutron elastic scattering and small cross sections for neutron-nucleus  interactions including absorption \cite{ebisawa1979neutron,hino2004recent,eriksson2023morphology}.
As shown in Fig.~\ref{F.NiElasC}, the elastic scattering cross sections have a constant value of approximately 20 barn in the thermal and epithermal neutron energy region from mev to KeV.  
\\
In this study, we design a tube collimator made of Ni-0 (natural)
to increase the collection efficiency of neutrons at a detector located on the end of a beamline.
Fig.~\ref{F.NiColli(a)} shows the schematic view of the tube collimation for neutrons.
If a neutron is emitted from a neutron source for a direction close to the center axis of a collimator, the neutron is expected to arrive directly at a detector.
However, when a neutron is radiated toward the wall of the collimator, the neutron collides on the wall and it may pass through the wall or scatter again toward the inside of the collimator.
%When a neutron incidents into the nickle surface, the length of the total flight of a neutron from the source to a detector depends on the incident angle (as $\theta$ in the Fig. \ref{F.NiColli(a)}). 
The probability of the elastic scattering depends on the incident angle $\theta$ in the Fig.~\ref{F.NiColli(a)}.
With increasing the incident angel, the probability of the scattering of the incident neutron decreases so that 
most incident neutrons pass through the wall toward the outside of the tube when the wall is thin. 
In contrast, with decreasing the incident angle, the probability of scattering toward the inside of the tube increases.
When a neutron scatters with a relatively large angle the neutron may again scatter on another part of the same tube.
This means that a neutron may arrive at the detector after multiple scattering on the wall.
The probability of scattering depends also on the thickness of the wall.
If a wall is thick enough to scatter almost all incident neutrons, a part of neutrons arrive at the detector through many times of scattering.
This causes the time broadening at the detector in the TOF method.
When the incident angle is small, the length of the total flight pass of the neutron becomes shorter than that in the case of the larger incident angle.
As a result, the delay of the arriving time at the detector becomes shorter and the finally obtained energy close to the energy in the case that the neutron reach directly on the detector without any scattering on the wall. 
In the case of multiple scattering, the total fight length becomes much longer than the direct flight length.
Therefore, it is desired that neutrons arrive directly at the detector or through a small number of multiple scattering.
Because the average number of scattering depends on the thickness of the wall,
the thickness can be optimized to increase the collection efficiency of neutrons at the detector with a acceptable time spread.
\\
We should also consider the length of the beamline and the effective surface area of the detector
to determine the optimized thickness of the Ni wall.
Fig.~\ref{F.NiColli(b)} shows the calculated reaction rate with a function of the angle of the incident neutrons for 2 mm thickness Ni wall.
The neutron energy is assumed as 5~eV because this energy is close to resonance energies in $^{109}$Ag and $^{181}$Ta as discussed later.
When neutrons incident vertically ($\theta=90^\circ$) on the Ni wall, the scattering probability is approximately 20\%. 
As the incident angle $\theta$ decreases, the scattering probability increases exponentially. 
With a 1.8 m length beamline and a detector surface size of $\phi=$2.54 cm, we conclude the optimized thickness is 2 mm to provide almost 100\% scattering rate at the incident angles lower than $\theta$=10$^\circ$.    
By this design, the epithermal neutrons could be effectively collected by the detector along the beamline.

\begin{figure}[htbp]
	\subfigbottomskip=0pt
	\subfigcapskip=0pt 
	\begin{center}
		\subfigure[\label{F.NiSimu(a)}]{
			\includegraphics[width=0.6\textwidth]{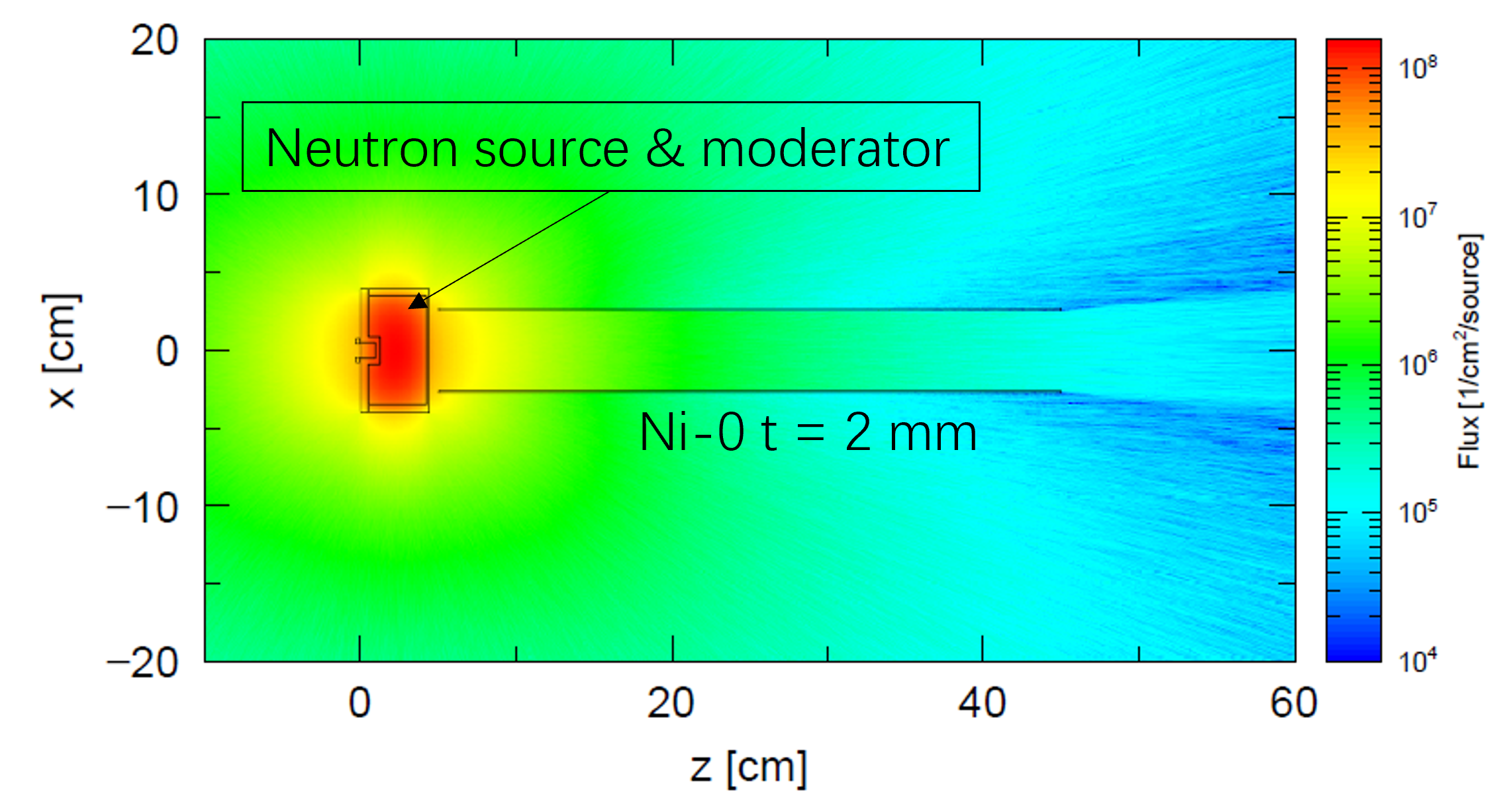}%
		}
		\subfigure[\label{F.NiSimu(b)}]{
			\includegraphics[width=0.4\textwidth]{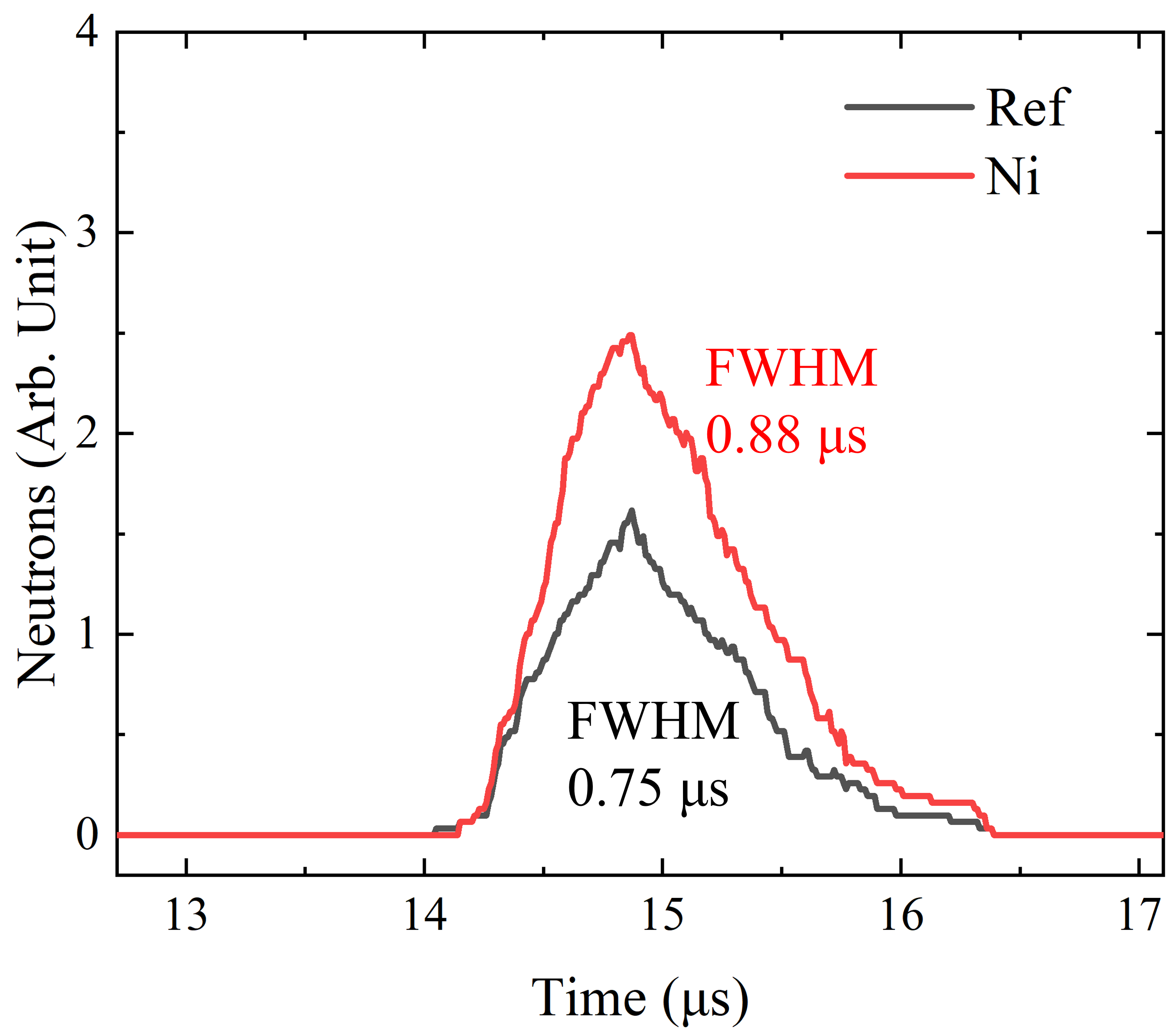}%
		}
	\end{center}
	\caption{\label{F.NiSimu}(a) The 2D result of simulated neutron collimation by using a nickle tube (L = 45 cm, t = 2 mm). The neutrons with 5 eV energy were recorded duration the simulation. (b) The time duration of 5 eV neutrons on the exit of nickle tube. The red line is the 5eV neutron pulse at exit of the nickle tube. The black line is a reference result with the same setup except the nickle tube.} 
\end{figure}

The neutron transport with the designed nickle tube was calculated using the PHITS Monte Carlo particle transport simulation code \cite{sato2018features}. 
We set the previously measured energy spectrum of the fast neutrons generated with LFEX laser as the input source. 
The neutrons with an energy of 5$\pm$0.05 eV were traced during the simulation. 
As shown in Fig.~\ref{F.NiSimu(a)}, a 45 cm length nickle tube with a thickness of 2~mm was placed at downstream of the neutron moderator with the source. 
The two-dimensional distribution of the neutrons shows that the 5 eV energy neutron flux inside the nickle tube is approximately twice of that in the reference simulation (without nickle tube). 
To evaluate the neuron pulse duration at the detector position, the neutron transport without the nickle tube was also calculated as a reference.
The energy window of $\pm$0.05 eV leads the broadening of the initial pulse duration at the distance of 45~cm between the neutron source and the detector. 
The neutron pulse duration at 5 eV in the reference is 0.75 $\mu$s in full width at half maximum (FWHM).
In the case with the nickle tube the neutron pulse duration is 0.88 $\mu$s in FWHM,
which is broader than the pulse width of the reference by 0.13 $\mu$s. 
This result shows that when we use the presently designed tube we could measure the TOF spectrum with the time resolution that is not drastically wider than that without the collimator.
Note that in the laboratory experiment the energy resolution depend also on the wide energy spread of the initial neutron pulse.
\subsection{The design of background shield}
We made the tube as design above for the experimental setup.
To decrease the background neutrons which are scattered by the laser chamber and other structures, a shield sleeve was set on the outside of the nickle tube (Fig.~\ref{F.NiExp}). The sleeve was made of polyethylene doped with 10\% B$_{2}$O$_{3}$. The high hydrogen content of polyethylene could decelerate the neutrons to low-energy thermal region. 
Boron is widely used as low-energy neutron absorber due to its large neutron capture cross-sections. 
The nickle tube installed on the laser chamber is folded by the aluminum (Al) guider because Al has very small neutron reaction cross-sections to provide a minimal interference on the neutron beamline. 
\begin{figure}[htbp]
	\subfigbottomskip=0pt
	\subfigcapskip=0pt 
	\begin{center}
		\includegraphics[width=0.4\textwidth]{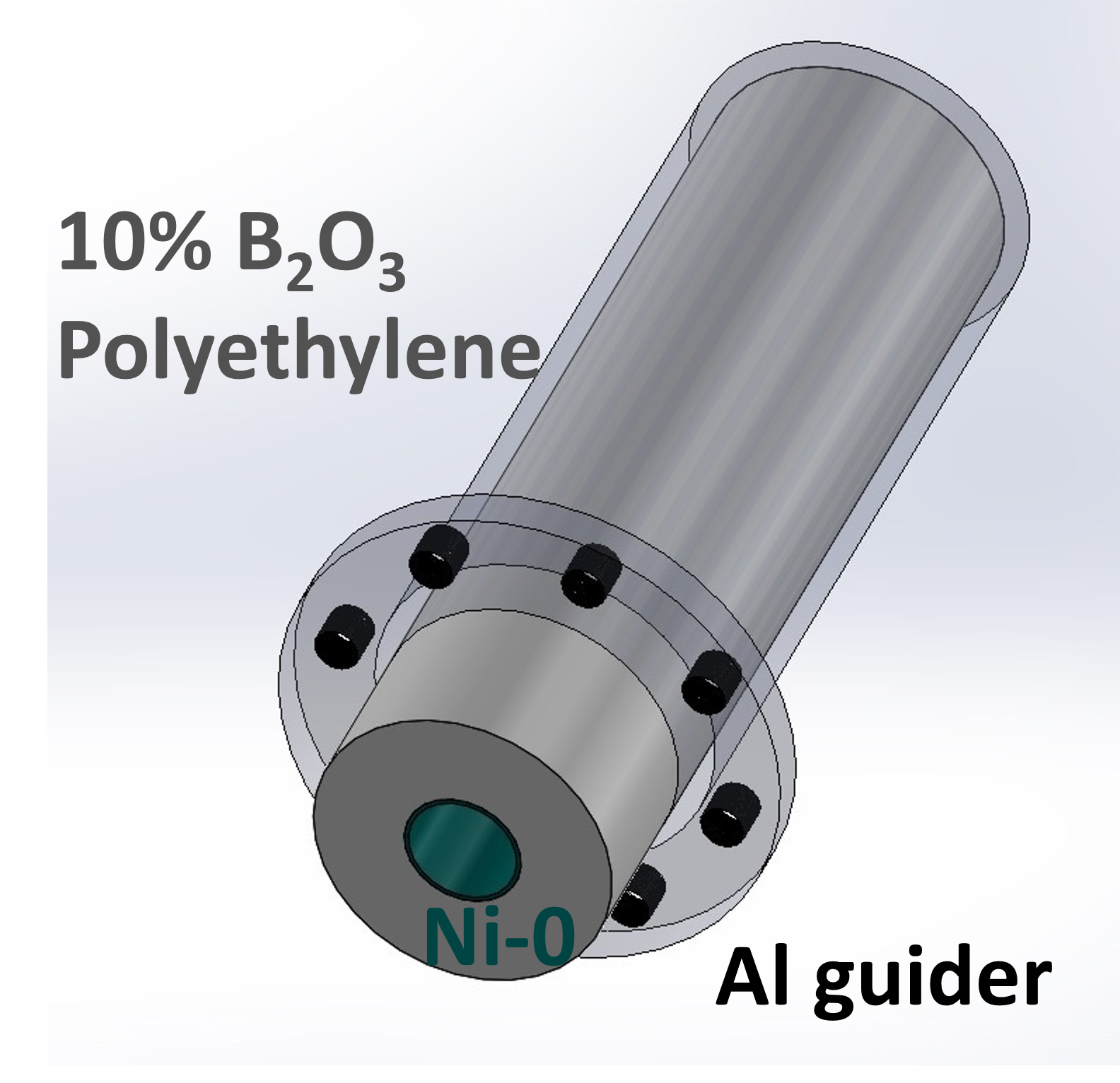}%
	\end{center}
	\caption{\label{F.NiBPoly} The structure diagram of neutron collimator used in LDNS experiments.}
\end{figure} 
\subsection{The experimental demonstration of the beamline development}
\begin{figure}[htbp]
	\subfigbottomskip=0pt
	\subfigcapskip=0pt 
	\begin{center}
		\includegraphics[width=0.6\textwidth]{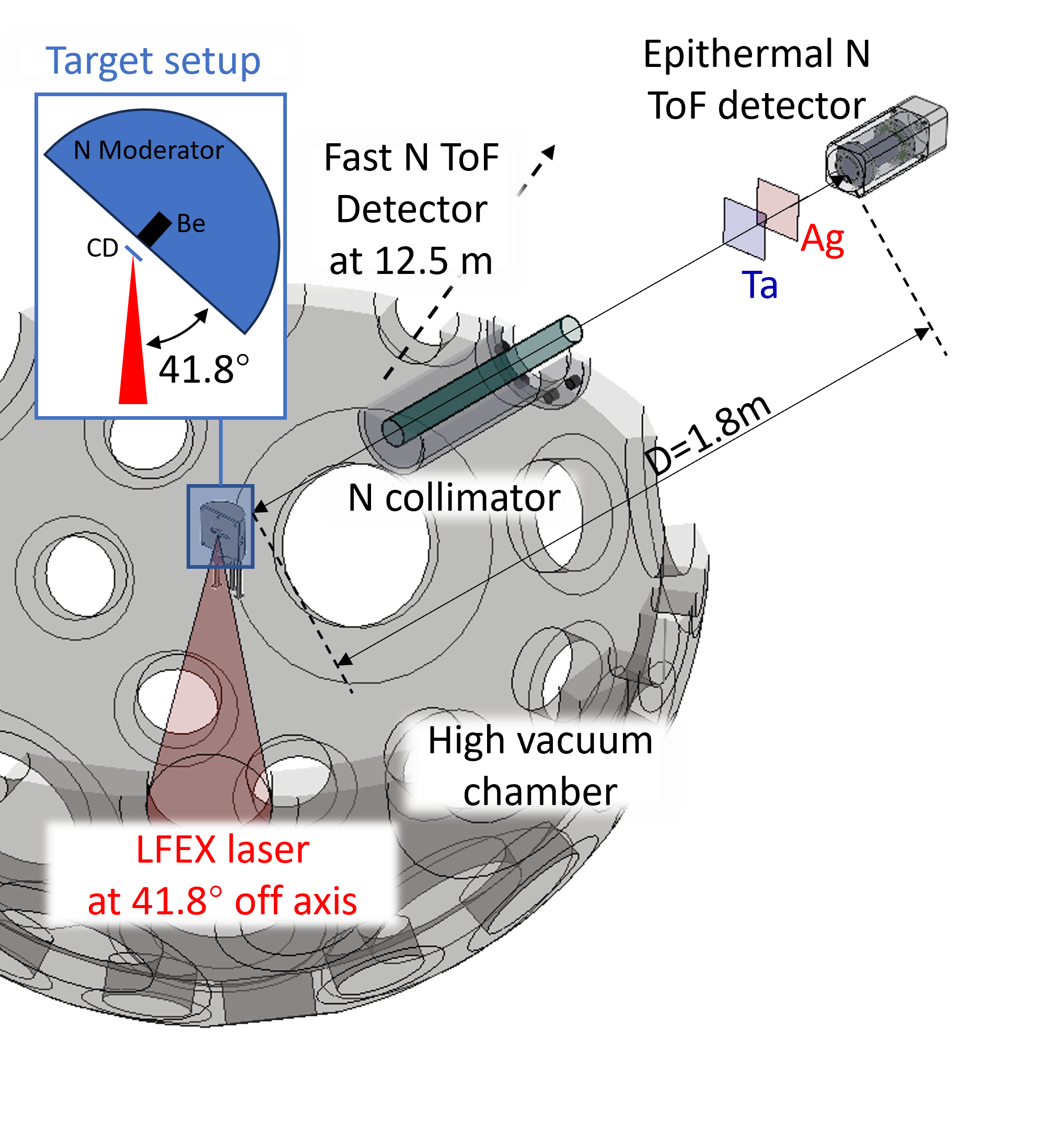}%
	\end{center}
	\caption{\label{F.EXPSetup} The structure diagram of neutron collimator used in LDNS experiments.}
\end{figure}

\begin{figure}[htbp]
	\subfigbottomskip=0pt
	\subfigcapskip=0pt 
	\begin{center}
		\subfigure[\label{F.NiExp(a)}]{
			\includegraphics[width=0.45\textwidth]{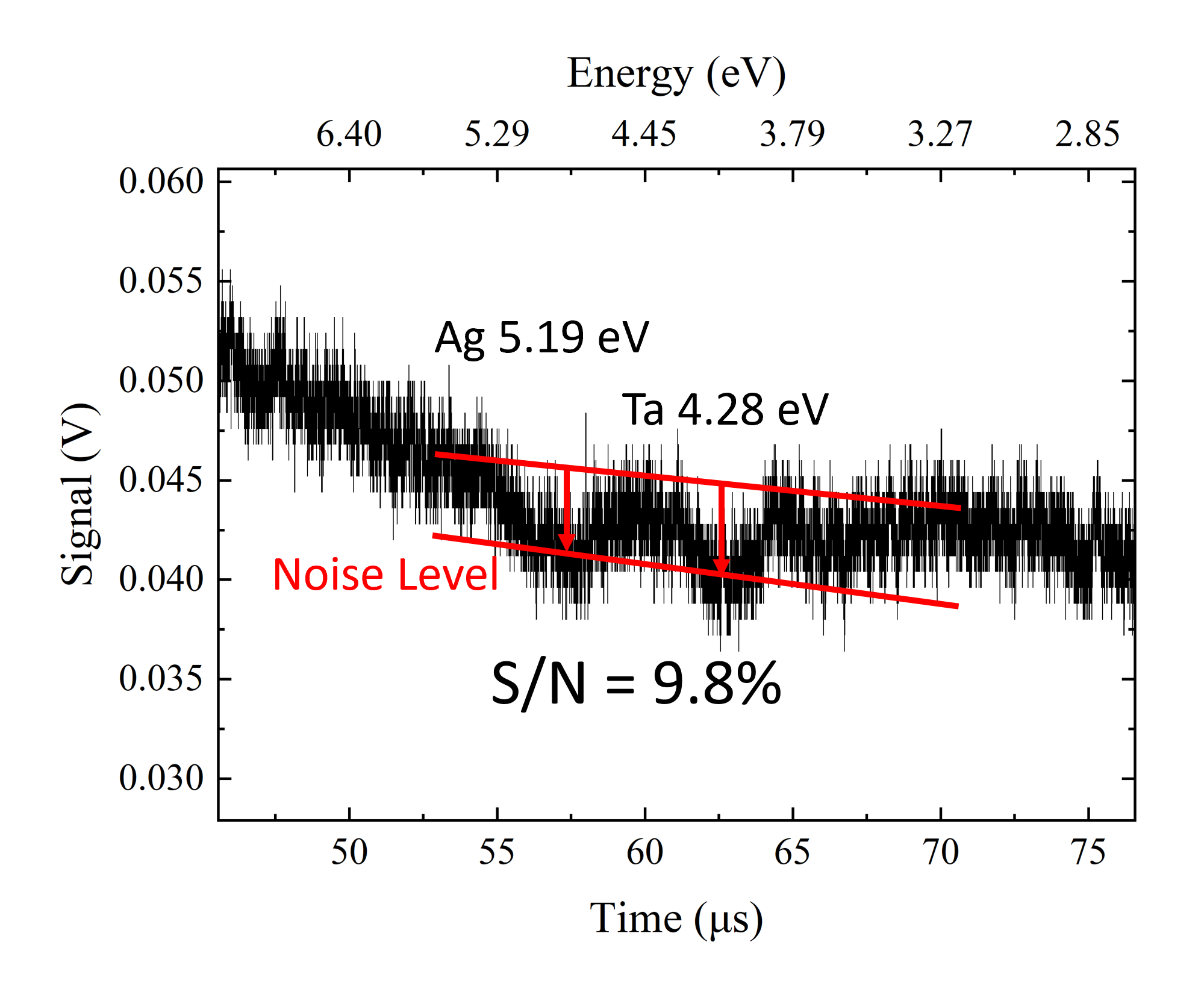}%
		}
		\subfigure[\label{F.NiExp(b)}]{
			\includegraphics[width=0.45\textwidth]{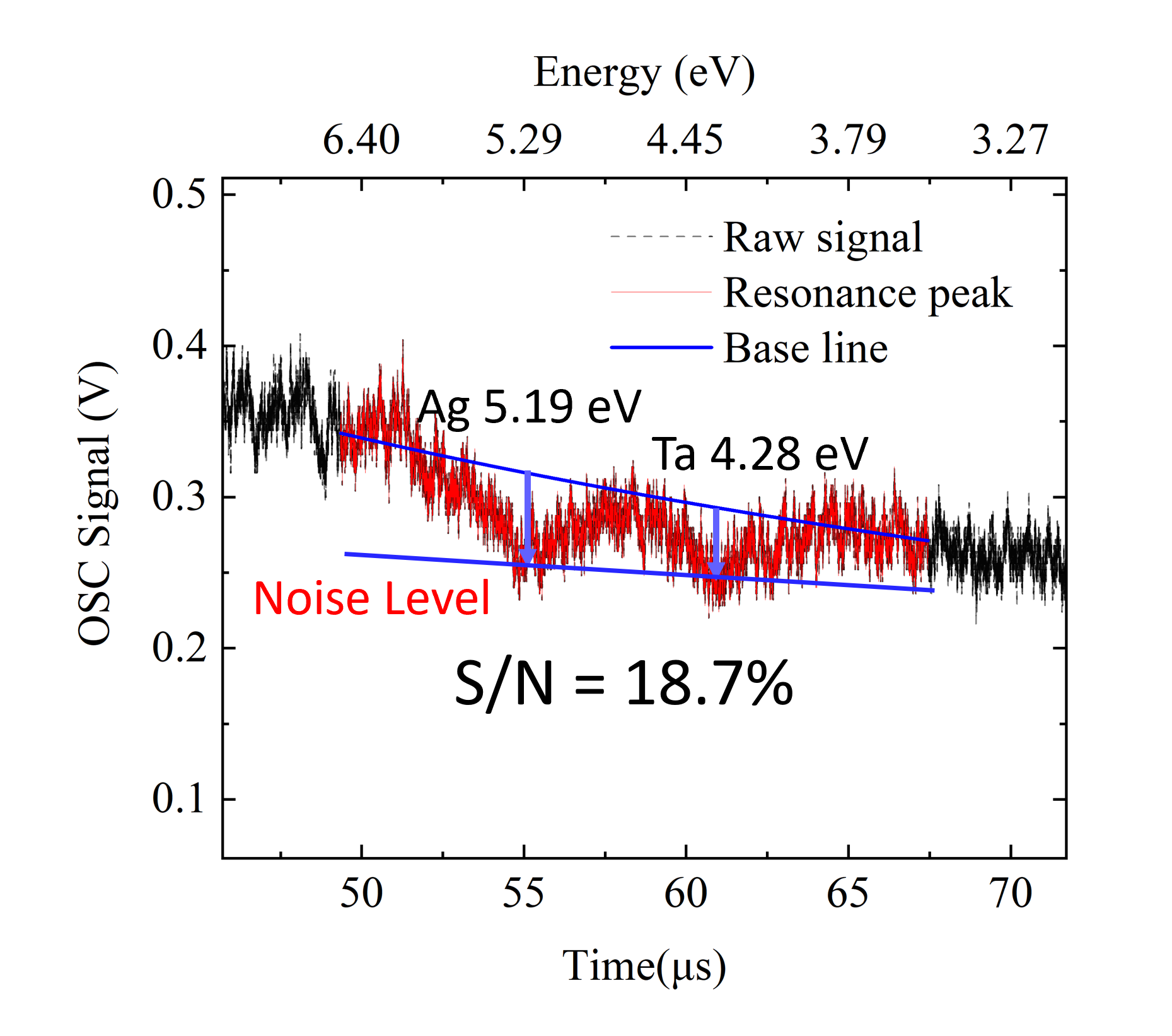}%
		}
	\end{center}
	\caption{\label{F.NiExp}(a) The experimental result of NRS using LDNS without any shield and collimator. The resonance samples are Ag and Ta. (b) The experimental result of NRS using LDNS with the nickle collimator design above. The resonance samples are Ag and Ta.} 
\end{figure}

We demonstrated the effect of the presently developed collimator using the nuclear resonance spectroscopy with LFEX laser \cite{miyanaga200610} at Institute for laser engineering (ILE) in Osaka University.
The experiment was carried out using 1.5 ps laser pulses and a total energy of 900 J on the target.
A laser pulse with an intensity of approximately 1$\times$10$^{19}$ W/cm$^{2}$ was focused on a 5 $\mu$m thickness deuterated polystyrene foil to accelerate protons and deuterons up to a few tens MeV.
A Be cylinder target with a diameter of 10 mm and a thickness of 10 mm was set behind the ion acceleration target.
The fast neutrons with energies of up to a few tens MeV are produced by the $^{9}$Be(p, n)$^{9}$B and $^{9}$Be(d, n)$^{10}$B nuclear reactions in the Be cylinder target,
and the fast neutrons were decelerated with a moderator around the Be target.
The experimental setup is shown in Fig. \ref{F.EXPSetup}. 
The energy spectrum of the fast neutrons were measured with a TOF method with a plastic scintillation detector and 
the thermal and epithermal neutrons were measured with  a $^{6}$Li-doped glass scintillator (GS20, Scintacor Ltd., 10 mm in thickness) coupled to a time-gated photomultiplier tube developed based on a HAMAMATSU-R2083.
The neutron collimator was set in the 1.8~m beamline. The entrance of the collimator is 48~cm far from the neutron source.
To measure neutron absorption resonances we placed two metallic foils of Ag and Ta with a thickness 0.2 and 0.1 mm, respectively, in the front of the detector on the beamline.
The resonance peaks of $^{109}$Ag at 5.19 eV and $^{181}$Ta at 4.28 eV were used for checking of neutron signal levels. 
\\
Fig. \ref{F.NiExp(a)} presents the TOF spectrum measured without the collimator, whereas Fig. \ref{F.NiExp(b)} shows the spectrum measured with the presently developed neutron collimator.
A same detector is used in the two cases with the same setup.
The thicknesses of the resonance samples were adjusted to absorb almost 100\% neutrons at the resonance energies. 
We define the depths at the resonance energies as the signal level.
We evaluated the signal-to-noise (S/N) ratio defined by the ratio of the signal to the noise level of the base line except for the neutron resonance absorption. The depth of each resonance is estimated by a fitted line of the spectrum. 
The signals in Fig. \ref{F.NiExp} indicate that the developed beamline with the neutron collimator improved the S/N ratio from 9.8\% to 18.7\%. 
This shows that the presently developed collimator is effective for the TOF measurement using LDNS.
\section{Conclusion}
We have developed a neutron beamline optimally designed for the laser-driven nuclear resonance spectroscopy.
The thin wall of the Ni-0 collimator allows neutrons having a smaller incident angle to be reflected by the wall and arrive at the detector, whereas neutrons having a larger incident angle penetrate the wall.
As a result, the collection efficiency of neutrons have been enhanced with only small broadening the temporal resolution of the TOF analysis, while the background of neutron scattering and X-rays were inhibited by the shield from entering the beamline. The collimator and shield lead to the improvement on the S/N ration by a factor of approximately two.
\\
The thickness and length of the nickle tube can be adjusted to selectively collimate neutrons in a desired energy region.
This result indicates that the neutron collimator and shield developed here is flexibly applicable to different type of analysis driven by LDNS.

\section{Data Availability Statement}
Data sets generated during the current study are available from the corresponding author on reasonable request.

\section{Acknowledgments}
This work was funded by Grant-in-Aid for Scientific Research (No. 25420911, No. 26246043, and
No. 22H02007) of MEXT, A-STEP (AS2721002c), and PRESTO(JPMJPR15PD) commissioned by JST. The authors
thank the technical support staff of ILE for their assistance with the laser operation, target fabrication, and plasma
diagnostics. This work was supported by the Collaboration Research Program of ILE, Osaka University.
\bibliography{sn-bibliography}% common bib file


\begin{thebibliography}{21}
\ifx \bisbn   \undefined \def \bisbn  #1{ISBN #1}\fi
\ifx \binits  \undefined \def \binits#1{#1}\fi
\ifx \bauthor  \undefined \def \bauthor#1{#1}\fi
\ifx \batitle  \undefined \def \batitle#1{#1}\fi
\ifx \bjtitle  \undefined \def \bjtitle#1{#1}\fi
\ifx \bvolume  \undefined \def \bvolume#1{\textbf{#1}}\fi
\ifx \byear  \undefined \def \byear#1{#1}\fi
\ifx \bissue  \undefined \def \bissue#1{#1}\fi
\ifx \bfpage  \undefined \def \bfpage#1{#1}\fi
\ifx \blpage  \undefined \def \blpage #1{#1}\fi
\ifx \burl  \undefined \def \burl#1{\textsf{#1}}\fi
\ifx \doiurl  \undefined \def \doiurl#1{\url{https://doi.org/#1}}\fi
\ifx \betal  \undefined \def \betal{\textit{et al.}}\fi
\ifx \binstitute  \undefined \def \binstitute#1{#1}\fi
\ifx \binstitutionaled  \undefined \def \binstitutionaled#1{#1}\fi
\ifx \bctitle  \undefined \def \bctitle#1{#1}\fi
\ifx \beditor  \undefined \def \beditor#1{#1}\fi
\ifx \bpublisher  \undefined \def \bpublisher#1{#1}\fi
\ifx \bbtitle  \undefined \def \bbtitle#1{#1}\fi
\ifx \bedition  \undefined \def \bedition#1{#1}\fi
\ifx \bseriesno  \undefined \def \bseriesno#1{#1}\fi
\ifx \blocation  \undefined \def \blocation#1{#1}\fi
\ifx \bsertitle  \undefined \def \bsertitle#1{#1}\fi
\ifx \bsnm \undefined \def \bsnm#1{#1}\fi
\ifx \bsuffix \undefined \def \bsuffix#1{#1}\fi
\ifx \bparticle \undefined \def \bparticle#1{#1}\fi
\ifx \barticle \undefined \def \barticle#1{#1}\fi
\bibcommenthead
\ifx \bconfdate \undefined \def \bconfdate #1{#1}\fi
\ifx \botherref \undefined \def \botherref #1{#1}\fi
\ifx \url \undefined \def \url#1{\textsf{#1}}\fi
\ifx \bchapter \undefined \def \bchapter#1{#1}\fi
\ifx \bbook \undefined \def \bbook#1{#1}\fi
\ifx \bcomment \undefined \def \bcomment#1{#1}\fi
\ifx \oauthor \undefined \def \oauthor#1{#1}\fi
\ifx \citeauthoryear \undefined \def \citeauthoryear#1{#1}\fi
\ifx \endbibitem  \undefined \def \endbibitem {}\fi
\ifx \bconflocation  \undefined \def \bconflocation#1{#1}\fi
\ifx \arxivurl  \undefined \def \arxivurl#1{\textsf{#1}}\fi
\csname PreBibitemsHook\endcsname

%%% 1
\bibitem[\protect\citeauthoryear{Roth et~al.}{2013}]{roth2013bright}
\begin{barticle}
\bauthor{\bsnm{Roth}, \binits{M.}},
\bauthor{\bsnm{Jung}, \binits{D.}},
\bauthor{\bsnm{Falk}, \binits{K.}},
\bauthor{\bsnm{Guler}, \binits{N.}},
\bauthor{\bsnm{Deppert}, \binits{O.}},
\bauthor{\bsnm{Devlin}, \binits{M.}},
\bauthor{\bsnm{Favalli}, \binits{A.}},
\bauthor{\bsnm{Fernandez}, \binits{J.}},
\bauthor{\bsnm{Gautier}, \binits{D.}},
\bauthor{\bsnm{Geissel}, \binits{M.}}, \betal:
\batitle{Bright laser-driven neutron source based on the relativistic
  transparency of solids}.
\bjtitle{Physical review letters}
\bvolume{110}(\bissue{4}),
\bfpage{044802}
(\byear{2013})
\end{barticle}
\endbibitem

%%% 2
\bibitem[\protect\citeauthoryear{Kar et~al.}{2016}]{kar2016beamed}
\begin{barticle}
\bauthor{\bsnm{Kar}, \binits{S.}},
\bauthor{\bsnm{Green}, \binits{A.}},
\bauthor{\bsnm{Ahmed}, \binits{H.}},
\bauthor{\bsnm{Alejo}, \binits{A.}},
\bauthor{\bsnm{Robinson}, \binits{A.}},
\bauthor{\bsnm{Cerchez}, \binits{M.}},
\bauthor{\bsnm{Clarke}, \binits{R.}},
\bauthor{\bsnm{Doria}, \binits{D.}},
\bauthor{\bsnm{Dorkings}, \binits{S.}},
\bauthor{\bsnm{Fernandez}, \binits{J.}}, \betal:
\batitle{Beamed neutron emission driven by laser accelerated light ions}.
\bjtitle{New Journal of Physics}
\bvolume{18}(\bissue{5}),
\bfpage{053002}
(\byear{2016})
\end{barticle}
\endbibitem

%%% 3
\bibitem[\protect\citeauthoryear{Alejo et~al.}{2017}]{alejo2017high}
\begin{barticle}
\bauthor{\bsnm{Alejo}, \binits{A.}},
\bauthor{\bsnm{Krygier}, \binits{A.}},
\bauthor{\bsnm{Ahmed}, \binits{H.}},
\bauthor{\bsnm{Morrison}, \binits{J.}},
\bauthor{\bsnm{Clarke}, \binits{R.}},
\bauthor{\bsnm{Fuchs}, \binits{J.}},
\bauthor{\bsnm{Green}, \binits{A.}},
\bauthor{\bsnm{Green}, \binits{J.}},
\bauthor{\bsnm{Jung}, \binits{D.}},
\bauthor{\bsnm{Kleinschmidt}, \binits{A.}}, \betal:
\batitle{High flux, beamed neutron sources employing deuteron-rich ion beams
  from d2o-ice layered targets}.
\bjtitle{Plasma Physics and Controlled Fusion}
\bvolume{59}(\bissue{6}),
\bfpage{064004}
(\byear{2017})
\end{barticle}
\endbibitem

%%% 4
\bibitem[\protect\citeauthoryear{Kleinschmidt
  et~al.}{2018}]{kleinschmidt2018intense}
\begin{barticle}
\bauthor{\bsnm{Kleinschmidt}, \binits{A.}},
\bauthor{\bsnm{Bagnoud}, \binits{V.}},
\bauthor{\bsnm{Deppert}, \binits{O.}},
\bauthor{\bsnm{Favalli}, \binits{A.}},
\bauthor{\bsnm{Frydrych}, \binits{S.}},
\bauthor{\bsnm{Hornung}, \binits{J.}},
\bauthor{\bsnm{Jahn}, \binits{D.}},
\bauthor{\bsnm{Schaumann}, \binits{G.}},
\bauthor{\bsnm{Tebartz}, \binits{A.}},
\bauthor{\bsnm{Wagner}, \binits{F.}}, \betal:
\batitle{Intense, directed neutron beams from a laser-driven neutron source at
  phelix}.
\bjtitle{Physics of Plasmas}
\bvolume{25}(\bissue{5}),
\bfpage{053101}
(\byear{2018})
\end{barticle}
\endbibitem

%%% 5
\bibitem[\protect\citeauthoryear{Zimmer et~al.}{2022}]{zimmer2022demonstration}
\begin{barticle}
\bauthor{\bsnm{Zimmer}, \binits{M.}},
\bauthor{\bsnm{Scheuren}, \binits{S.}},
\bauthor{\bsnm{Kleinschmidt}, \binits{A.}},
\bauthor{\bsnm{Mitura}, \binits{N.}},
\bauthor{\bsnm{Tebartz}, \binits{A.}},
\bauthor{\bsnm{Schaumann}, \binits{G.}},
\bauthor{\bsnm{Abel}, \binits{T.}},
\bauthor{\bsnm{Ebert}, \binits{T.}},
\bauthor{\bsnm{Hesse}, \binits{M.}},
\bauthor{\bsnm{Z{\"a}hter}, \binits{{\c{S}}.}}, \betal:
\batitle{Demonstration of non-destructive and isotope-sensitive material
  analysis using a short-pulsed laser-driven epi-thermal neutron source}.
\bjtitle{Nature Communications}
\bvolume{13}(\bissue{1}),
\bfpage{1173}
(\byear{2022})
\end{barticle}
\endbibitem

%%% 6
\bibitem[\protect\citeauthoryear{Yogo et~al.}{2023}]{yogo2023advances}
\begin{barticle}
\bauthor{\bsnm{Yogo}, \binits{A.}},
\bauthor{\bsnm{Arikawa}, \binits{Y.}},
\bauthor{\bsnm{Abe}, \binits{Y.}},
\bauthor{\bsnm{Mirfayzi}, \binits{S.}},
\bauthor{\bsnm{Hayakawa}, \binits{T.}},
\bauthor{\bsnm{Mima}, \binits{K.}},
\bauthor{\bsnm{Kodama}, \binits{R.}}:
\batitle{Advances in laser-driven neutron sources and applications}.
\bjtitle{The European Physical Journal A}
\bvolume{59}(\bissue{8}),
\bfpage{191}
(\byear{2023})
\end{barticle}
\endbibitem

%%% 7
\bibitem[\protect\citeauthoryear{Arikawa
  et~al.}{2023}]{arikawa2023demonstration}
\begin{barticle}
\bauthor{\bsnm{Arikawa}, \binits{Y.}},
\bauthor{\bsnm{Morace}, \binits{A.}},
\bauthor{\bsnm{Abe}, \binits{Y.}},
\bauthor{\bsnm{Iwata}, \binits{N.}},
\bauthor{\bsnm{Sentoku}, \binits{Y.}},
\bauthor{\bsnm{Yogo}, \binits{A.}},
\bauthor{\bsnm{Matsuo}, \binits{K.}},
\bauthor{\bsnm{Nakai}, \binits{M.}},
\bauthor{\bsnm{Nagatomo}, \binits{H.}},
\bauthor{\bsnm{Mima}, \binits{K.}}, \betal:
\batitle{Demonstration of efficient relativistic electron acceleration by
  surface plasmonics with sequential target processing using high repetition
  lasers}.
\bjtitle{Physical Review Research}
\bvolume{5}(\bissue{1}),
\bfpage{013062}
(\byear{2023})
\end{barticle}
\endbibitem

%%% 8
\bibitem[\protect\citeauthoryear{G{\"u}nther et~al.}{2022}]{gunther2022forward}
\begin{barticle}
\bauthor{\bsnm{G{\"u}nther}, \binits{M.}},
\bauthor{\bsnm{Rosmej}, \binits{O.}},
\bauthor{\bsnm{Tavana}, \binits{P.}},
\bauthor{\bsnm{Gyrdymov}, \binits{M.}},
\bauthor{\bsnm{Skobliakov}, \binits{A.}},
\bauthor{\bsnm{Kantsyrev}, \binits{A.}},
\bauthor{\bsnm{Z{\"a}hter}, \binits{S.}},
\bauthor{\bsnm{Borisenko}, \binits{N.}},
\bauthor{\bsnm{Pukhov}, \binits{A.}},
\bauthor{\bsnm{Andreev}, \binits{N.}}:
\batitle{Forward-looking insights in laser-generated ultra-intense $\gamma$-ray
  and neutron sources for nuclear application and science}.
\bjtitle{Nature Communications}
\bvolume{13}(\bissue{1}),
\bfpage{170}
(\byear{2022})
\end{barticle}
\endbibitem

%%% 9
\bibitem[\protect\citeauthoryear{Mori et~al.}{2023}]{mori2023feasibility}
\begin{barticle}
\bauthor{\bsnm{Mori}, \binits{T.}},
\bauthor{\bsnm{Yogo}, \binits{A.}},
\bauthor{\bsnm{Arikawa}, \binits{Y.}},
\bauthor{\bsnm{Hayakawa}, \binits{T.}},
\bauthor{\bsnm{Mirfayzi}, \binits{S.R.}},
\bauthor{\bsnm{Lan}, \binits{Z.}},
\bauthor{\bsnm{Wei}, \binits{T.}},
\bauthor{\bsnm{Abe}, \binits{Y.}},
\bauthor{\bsnm{Nakai}, \binits{M.}},
\bauthor{\bsnm{Mima}, \binits{K.}}, \betal:
\batitle{Feasibility study of laser-driven neutron sources for pharmaceutical
  applications}.
\bjtitle{High Power Laser Science and Engineering}
\bvolume{11},
\bfpage{20}
(\byear{2023})
\end{barticle}
\endbibitem

%%% 10
\bibitem[\protect\citeauthoryear{Breit and Wigner}{1936}]{breit1936capture}
\begin{barticle}
\bauthor{\bsnm{Breit}, \binits{G.}},
\bauthor{\bsnm{Wigner}, \binits{E.}}:
\batitle{Capture of slow neutrons}.
\bjtitle{Physical review}
\bvolume{49}(\bissue{7}),
\bfpage{519}
(\byear{1936})
\end{barticle}
\endbibitem

%%% 11
\bibitem[\protect\citeauthoryear{Mirfayzi et~al.}{2020}]{mirfayzi2020miniature}
\begin{barticle}
\bauthor{\bsnm{Mirfayzi}, \binits{S.R.}},
\bauthor{\bsnm{Ahmed}, \binits{H.}},
\bauthor{\bsnm{Doria}, \binits{D.}},
\bauthor{\bsnm{Alejo}, \binits{A.}},
\bauthor{\bsnm{Ansell}, \binits{S.}},
\bauthor{\bsnm{Clarke}, \binits{R.}},
\bauthor{\bsnm{Gonzalez-Izquierdo}, \binits{B.}},
\bauthor{\bsnm{Hadjisolomou}, \binits{P.}},
\bauthor{\bsnm{Heathcote}, \binits{R.}},
\bauthor{\bsnm{Hodge}, \binits{T.}}, \betal:
\batitle{A miniature thermal neutron source using high power lasers}.
\bjtitle{Applied Physics Letters}
\bvolume{116}(\bissue{17}),
\bfpage{174102}
(\byear{2020})
\end{barticle}
\endbibitem

%%% 12
\bibitem[\protect\citeauthoryear{Mirfayzi et~al.}{2019}]{mirfayzi2019ultra}
\begin{botherref}
\oauthor{\bsnm{Mirfayzi}, \binits{S.}},
\oauthor{\bsnm{Mori}, \binits{T.}},
\oauthor{\bsnm{Rusby}, \binits{D.}},
\oauthor{\bsnm{Armstrong}, \binits{C.}},
\oauthor{\bsnm{Ahmed}, \binits{H.}},
\oauthor{\bsnm{Borghesi}, \binits{M.}},
\oauthor{\bsnm{Brenner}, \binits{C.}},
\oauthor{\bsnm{Clarke}, \binits{R.}},
\oauthor{\bsnm{Davidson}, \binits{Z.}},
\oauthor{\bsnm{Doria}, \binits{D.}}, et al.:
Ultra-short, beamed source of laser-driven epithermal neutrons.
Applied Physics Letters
(2019)
\end{botherref}
\endbibitem

%%% 13
\bibitem[\protect\citeauthoryear{Tremsin et~al.}{2014}]{tremsin2014neutron}
\begin{barticle}
\bauthor{\bsnm{Tremsin}, \binits{A.}},
\bauthor{\bsnm{Shinohara}, \binits{T.}},
\bauthor{\bsnm{Kai}, \binits{T.}},
\bauthor{\bsnm{Ooi}, \binits{M.}},
\bauthor{\bsnm{Kamiyama}, \binits{T.}},
\bauthor{\bsnm{Kiyanagi}, \binits{Y.}},
\bauthor{\bsnm{Shiota}, \binits{Y.}},
\bauthor{\bsnm{McPhate}, \binits{J.}},
\bauthor{\bsnm{Vallerga}, \binits{J.}},
\bauthor{\bsnm{Siegmund}, \binits{O.}}, \betal:
\batitle{Neutron resonance transmission spectroscopy with high spatial and
  energy resolution at the j-parc pulsed neutron source}.
\bjtitle{Nuclear Instruments and Methods in Physics Research Section A:
  Accelerators, Spectrometers, Detectors and Associated Equipment}
\bvolume{746},
\bfpage{47}--\blpage{58}
(\byear{2014})
\end{barticle}
\endbibitem

%%% 14
\bibitem[\protect\citeauthoryear{Schillebeeckx
  et~al.}{2012}]{schillebeeckx2012neutron}
\begin{barticle}
\bauthor{\bsnm{Schillebeeckx}, \binits{P.}},
\bauthor{\bsnm{Borella}, \binits{A.}},
\bauthor{\bsnm{Emiliani}, \binits{F.}},
\bauthor{\bsnm{Gorini}, \binits{G.}},
\bauthor{\bsnm{Kockelmann}, \binits{W.}},
\bauthor{\bsnm{Kopecky}, \binits{S.}},
\bauthor{\bsnm{Lampoudis}, \binits{C.}},
\bauthor{\bsnm{Moxon}, \binits{M.}},
\bauthor{\bsnm{Cippo}, \binits{E.P.}},
\bauthor{\bsnm{Postma}, \binits{H.}}, \betal:
\batitle{Neutron resonance spectroscopy for the characterization of materials
  and objects}.
\bjtitle{Journal of Instrumentation}
\bvolume{7}(\bissue{03}),
\bfpage{03009}
(\byear{2012})
\end{barticle}
\endbibitem

%%% 15
\bibitem[\protect\citeauthoryear{Yogo et~al.}{2023}]{yogo2023laser}
\begin{barticle}
\bauthor{\bsnm{Yogo}, \binits{A.}},
\bauthor{\bsnm{Lan}, \binits{Z.}},
\bauthor{\bsnm{Arikawa}, \binits{Y.}},
\bauthor{\bsnm{Abe}, \binits{Y.}},
\bauthor{\bsnm{Mirfayzi}, \binits{S.}},
\bauthor{\bsnm{Wei}, \binits{T.}},
\bauthor{\bsnm{Mori}, \binits{T.}},
\bauthor{\bsnm{Golovin}, \binits{D.}},
\bauthor{\bsnm{Hayakawa}, \binits{T.}},
\bauthor{\bsnm{Iwata}, \binits{N.}}, \betal:
\batitle{Laser-driven neutron generation realizing single-shot resonance
  spectroscopy}.
\bjtitle{Physical Review X}
\bvolume{13}(\bissue{1}),
\bfpage{011011}
(\byear{2023})
\end{barticle}
\endbibitem

%%% 16
\bibitem[\protect\citeauthoryear{Ebisawa et~al.}{1979}]{ebisawa1979neutron}
\begin{barticle}
\bauthor{\bsnm{Ebisawa}, \binits{T.}},
\bauthor{\bsnm{Achiwa}, \binits{N.}},
\bauthor{\bsnm{Yamada}, \binits{S.}},
\bauthor{\bsnm{Akiyoshi}, \binits{T.}},
\bauthor{\bsnm{Okamoto}, \binits{S.}}:
\batitle{Neutron reflectivities of ni-mn and ni-ti multilayers for
  monochromators and supermirrors}.
\bjtitle{Journal of Nuclear Science and Technology}
\bvolume{16}(\bissue{9}),
\bfpage{647}--\blpage{659}
(\byear{1979})
\end{barticle}
\endbibitem

%%% 17
\bibitem[\protect\citeauthoryear{Hino et~al.}{2004}]{hino2004recent}
\begin{barticle}
\bauthor{\bsnm{Hino}, \binits{M.}},
\bauthor{\bsnm{Sunohara}, \binits{H.}},
\bauthor{\bsnm{Yoshimura}, \binits{Y.}},
\bauthor{\bsnm{Maruyama}, \binits{R.}},
\bauthor{\bsnm{Tasaki}, \binits{S.}},
\bauthor{\bsnm{Yoshino}, \binits{H.}},
\bauthor{\bsnm{Kawabata}, \binits{Y.}}:
\batitle{Recent development of multilayer neutron mirror at kurri}.
\bjtitle{Nuclear Instruments and Methods in Physics Research Section A:
  Accelerators, Spectrometers, Detectors and Associated Equipment}
\bvolume{529}(\bissue{1-3}),
\bfpage{54}--\blpage{58}
(\byear{2004})
\end{barticle}
\endbibitem

%%% 18
\bibitem[\protect\citeauthoryear{Eriksson
  et~al.}{2023}]{eriksson2023morphology}
\begin{barticle}
\bauthor{\bsnm{Eriksson}, \binits{F.}},
\bauthor{\bsnm{Ghafoor}, \binits{N.}},
\bauthor{\bsnm{Broekhuijsen}, \binits{S.}},
\bauthor{\bsnm{Greczynski}, \binits{G.}},
\bauthor{\bsnm{Schell}, \binits{N.}},
\bauthor{\bsnm{Birch}, \binits{J.}}:
\batitle{Morphology control in ni/ti multilayer neutron mirrors by ion-assisted
  interface engineering and b 4 c incorporation}.
\bjtitle{Optical Materials Express}
\bvolume{13}(\bissue{5}),
\bfpage{1424}--\blpage{1439}
(\byear{2023})
\end{barticle}
\endbibitem

%%% 19
\bibitem[\protect\citeauthoryear{Shibata et~al.}{2011}]{shibata2011jendl}
\begin{barticle}
\bauthor{\bsnm{Shibata}, \binits{K.}},
\bauthor{\bsnm{Iwamoto}, \binits{O.}},
\bauthor{\bsnm{Nakagawa}, \binits{T.}},
\bauthor{\bsnm{Iwamoto}, \binits{N.}},
\bauthor{\bsnm{Ichihara}, \binits{A.}},
\bauthor{\bsnm{Kunieda}, \binits{S.}},
\bauthor{\bsnm{Chiba}, \binits{S.}},
\bauthor{\bsnm{Furutaka}, \binits{K.}},
\bauthor{\bsnm{Otuka}, \binits{N.}},
\bauthor{\bsnm{Ohsawa}, \binits{T.}}, \betal:
\batitle{Jendl-4.0: a new library for nuclear science and engineering}.
\bjtitle{Journal of Nuclear Science and Technology}
\bvolume{48}(\bissue{1}),
\bfpage{1}--\blpage{30}
(\byear{2011})
\end{barticle}
\endbibitem

%%% 20
\bibitem[\protect\citeauthoryear{Sato et~al.}{2018}]{sato2018features}
\begin{barticle}
\bauthor{\bsnm{Sato}, \binits{T.}},
\bauthor{\bsnm{Iwamoto}, \binits{Y.}},
\bauthor{\bsnm{Hashimoto}, \binits{S.}},
\bauthor{\bsnm{Ogawa}, \binits{T.}},
\bauthor{\bsnm{Furuta}, \binits{T.}},
\bauthor{\bsnm{Abe}, \binits{S.-i.}},
\bauthor{\bsnm{Kai}, \binits{T.}},
\bauthor{\bsnm{Tsai}, \binits{P.-E.}},
\bauthor{\bsnm{Matsuda}, \binits{N.}},
\bauthor{\bsnm{Iwase}, \binits{H.}}, \betal:
\batitle{Features of particle and heavy ion transport code system (phits)
  version 3.02}.
\bjtitle{Journal of Nuclear Science and Technology}
\bvolume{55}(\bissue{6}),
\bfpage{684}--\blpage{690}
(\byear{2018})
\end{barticle}
\endbibitem

%%% 21
\bibitem[\protect\citeauthoryear{Miyanaga et~al.}{2006}]{miyanaga200610}
\begin{bchapter}
\bauthor{\bsnm{Miyanaga}, \binits{N.}},
\bauthor{\bsnm{Azechi}, \binits{H.}},
\bauthor{\bsnm{Tanaka}, \binits{K.}},
\bauthor{\bsnm{Kanabe}, \binits{T.}},
\bauthor{\bsnm{Jitsuno}, \binits{T.}},
\bauthor{\bsnm{Kawanaka}, \binits{J.}},
\bauthor{\bsnm{Fujimoto}, \binits{Y.}},
\bauthor{\bsnm{Kodama}, \binits{R.}},
\bauthor{\bsnm{Shiraga}, \binits{H.}},
\bauthor{\bsnm{Knodo}, \binits{K.}}, \betal:
\bctitle{10-kj pw laser for the firex-i program}.
In: \bbtitle{Journal de Physique IV (Proceedings)},
vol. \bseriesno{133},
pp. \bfpage{81}--\blpage{87}
(\byear{2006}).
\bcomment{EDP sciences}
\end{bchapter}
\endbibitem

\end{thebibliography}

\end{document}